\documentclass[12pt,preprint]{aastex}

\begin{document}

\def\hal{${H\alpha}$ }
\def\na{\ion{Na}{1}\ }
\def\pot{\ion{K}{1}\ }
\def\li{\ion{Li}{1}\ }
\def\ca{\ion{Ca}{2}\ }
\def\kms{km s$^{-1}$\ }
\def\mj{M$_J$\ }
\def\msun{M$_\odot$\ }
\def\rsun{R$_\odot$\ }
\def\lsun{L$_\odot$\ }
\def\mj{M$_J$\ }
\def\teff{T$_{e\! f\! f}$~}
\def\gv{{\it g}~}
\def\vsini{{\it v}~sin{\it i}~}
\def\vrad{v$_{\it rad}$~}
\def\lbol{L$_{bol}$~}
\def\lhal{L$_{\H\alpha}$ }
\def\eqwhal{EW$_{H\alpha}$ }
\def\logten{$log_{10}$ }
\def\jh{$J$-$H$ }
\def\jk{$J$-$K$ }
\def\jks{$J$-$K_S$ }
\def\av{A$_V$ }
\def\aj{A$_J$ }
\def\aks{A$_{Ks}$ }
\def\bcj{BC$_J$ }

\title{Flared Disks and Silicate Emission in Young Brown Dwarfs}
\author{Subhanjoy Mohanty\altaffilmark{1}, Ray Jayawardhana\altaffilmark{2},
Antonella Natta\altaffilmark{3},\\ Takuya Fujiyoshi\altaffilmark{4},
Motohide Tamura\altaffilmark{5}, David Barrado y
Navascu\'es\altaffilmark{6}}
\altaffiltext{1}{Harvard-Smithsonian CfA, 60 Garden Street, Cambridge, MA
02138, U.S.A.  smohanty@cfa.harvard.edu}
\altaffiltext{2}{Astronomy Department, University of Michigan, 830 Dennison
Building, Ann Arbor, MI 48109, U.S.A.}
\altaffiltext{3}{Osservatorio Astrofisico di Arcetri, INAF, Largo E. Fermi
5, I-50125 Firenze, Italy.}
\altaffiltext{4}{Subaru Telescope, National Astronomical Observatory of
Japan, 650 North A'ohoku Place, Hilo, Hawaii.}
\altaffiltext{5}{National Astronomical Observatory of Japan, Osawa 2-21-1,
Mitaka, Tokyo 181-8588, Japan.}
\altaffiltext{6}{Laboratorio de Astrof\'{\i}sica Espacial y F\'{\i}sica
Fundamental, INTA, P.O. Box 50727, E-2808 Madrid, Spain.}

\begin{abstract}
We present mid-infrared photometry of three very young brown dwarfs located
in the $\rho$ Ophiuchi star-forming region -- GY5, GY11 and GY310 --
obtained with the Subaru 8-meter telescope. All three sources were detected
at 8.6 and 11.7$\mu$m, confirming the presence of significant mid-infrared
excess arising from optically thick dusty disks.  The spectral energy
distributions of both GY310 and GY11 exhibit strong evidence of flared
disks; flat disks can be ruled out for these two brown dwarfs.  The data for
GY5 show large scatter, and are marginally consistent with both flared and
flat configurations.  Inner holes a few substellar radii in size are
indicated in all three cases (and especially in GY11), in agreement with
magnetospheric accretion models.  Finally, our 9.7$\mu$m flux for GY310
implies silicate emission from small grains on the disk surface (though the
data do not completely preclude larger grains with no silicate feature).
Our results demonstrate that disks around young substellar objects are
analogous to those girdling classical T Tauri stars, and exhibit a similar
range of disk geometries and dust properties.
\end{abstract}


\section{Introduction}
Now that a large number of brown dwarfs are known, some with inferred masses
close to those of extrasolar giant planets, their origin has become a
subject of widespread interest and fundamental importance.  One key recent
finding is that substellar objects harbor accretion disks and undergo a T
Tauri phase similar to that of low-mass stars, possibly suggesting a common
formation mechanism.  The evidence for disks around brown dwarfs includes
near- and mid-infrared excess (e.g., Tamura et al. 1998; Natta \& Testi
2001; Natta et al. 2002; Jayawardhana et al. 2003), spectroscopic signatures
of accretion (e.g., Jayawardhana, Mohanty \& Basri 2002, 2003; Muzerolle et
al. 2003) and millimeter-wave emission (Klein et al. 2003). In some cases,
these disks appear to last for up to 10 million years (Mohanty, Jayawardhana
\& Barrado y Navascu\'es 2003), a timescale comparable to that of disks
surrounding low-mass stars.  Moreover, there are now indications of
jets/outflows from objects close to or below the substellar limit
(Fern\'andez and Comer\'on 2001; Barrado y Navascu\'es, Mohanty \&
Jayawardhana 2004), further strengthening the T Tauri analogy.

Given the much lower mass and intrinsic luminosity of brown dwarfs relative
to T Tauri stars, the preceding discoveries raise intriguing questions about
how the detailed properties of brown dwarf disks compare to those of T Tauri
disks.  Observations in the mid-infrared, where optically thick disk
emission dominates over the substellar photosphere, can provide important
constraints on the disk models.  Here we report ground-based photometry in
the 8.6, 9.7 and 11.7$\mu$m filters, spanning the 10$\mu$m silicate feature,
of three young brown dwarfs in the nearby $\rho$ Ophiuchi region
(age$\lesssim$1 Myr, d$\approx$150 pc), and use these data to explore their
disk characteristics in comparison with models.

\section{Observations and Data Reduction}
All our photometry was obtained with the Cooled Mid-Infrared Camera and
Spectrograph (COMICS; Okamoto et al. 2003) on the Subaru telescope, on Mauna
Kea.  In imaging mode, the instrument offers a
42\arcsec$\times$32\arcsec$\,$ field of view, with a plate scale of
0.13\arcsec/pixel (diffraction-limited seeing at $\sim$10$\mu$m is
0.31\arcsec, or $\sim$2pix); we observed under seeing conditions of 1\arcsec 
or
better.  We acquired data through 3 filters: a narrow-band filter centered
on 8.6$\mu$m (width $\approx$ 0.4$\mu$m), and two broad-band filters
centered on 9.7 and 11.7$\mu$m (widths $\approx$ 1$\mu$m).  [GY92]5 (ISO-Oph
30) and [GY92]310 (ISO-Oph 164) were observed on 2003 June 18 UT; [GY92]11
(ISO-Oph 33) was observed on 2003 July 16 UT.  The flux standards HR5013 and
HD152880 were also observed on the former night, and HD152880 alone on the
latter.  Data were obtained in standard chop-and-nod mode for science
targets and calibrators. Time constraints only allowed observations at 8.6
and 11.7$\mu$m for GY11, and not at 9.7$\mu$m; the other two $\rho$ Oph 
sources were observed in all three filters.  Typical integration times were 
$\sim$1200s
per filter for the science targets.

Aperture photometry was performed on the dark-subtracted, flat-fielded and
coadded final images, using the {\it apphot} package in {\it
IRAF}\footnote{{\it IRAF} is distributed by the National Optical Astronomy
Observatories, operated by the Association of Universities for Research in
Astronomy, Inc., under contract to the National Science Foundation, USA.}.
Our final errors are $\lesssim$ 0.05 mag in calibration for HR5013 and
HD152880, and $\sim$0.15 mag in aperture photometry for GY5 and GY310,
resulting in a total error of $\sim$15\% in the inferred flux in all filters
for the latter two objects.  For the faintest source GY11, we conservatively
estimate an uncertainty of $\sim$0.25 mag in our photometry, yielding a
final flux error of $\sim$25\% at both 8.6 and 11.7$\mu$m.

GY5 is observed but not detected at 9.7$\mu$m, in either the individual nod
frames or in the coadded frames at a single nod position. Under the
circumstances, we derive a 3$\sigma$ detection upper limit of 35 mJy from
the individual frames.  This value is consistent with our detected 9.7$\mu$m
flux of $\sim$35 mJy in GY310, observed for nearly the same length of time:
the latter source is just visible at the 2--3$\sigma$ level in the
individual nod frames (but clearly detected at $\sim$6$\sigma$ in the
co-added frames). Our Subaru fluxes are listed in Table 1. In our analysis
of the spectral energy distributions (SEDs), we have also included the ISO
6.7 and 14.3 $\mu$m fluxes derived for these sources by Bontemps et al.
(2001) with an angular resolution of 6 arcsec, and the ISO 3.6, 4.5 and 6.0
$\mu$m fluxes for GY5 and 11 from Comeron et al. (1998) with 3-arcsec
resolution; these all have errors $\geq$ 10\%.

\section{Stellar Parameters}
We are interested here in the luminosity, mass and radius of our targets.
Mass is pertinent for establishing whether these are bona-fide brown dwarfs,
while luminosity and (to a lesser extent) mass and radius are important for
modelling their disk properties (\S 4).  Our stellar parameters for GY5, 11
and 310 are adopted from Natta et al. (2002, hereafter NTC02); the reader is
referred to that work for a detailed discussion of the methodology.  The
derived quantities are listed in Table 1.  NTC02 estimate errors of
$\pm$100K in \teff, 20--30\% in \lbol, and 20--30\% in mass (the \lbol
errors derive mainly from those in \av, and are likely somewhat higher in
the faintest, heavily reddened objects; see discussion of GY11 in \S 5).
They find a spectral type of M6, \teff of 2700K and mass of 60$\pm$20 \mj
for both GY5 and 310, and M8.5, 2400K and 10$\pm$2 \mj for GY11.  The
inferred bolometric luminosities for GY5 and 310 are very similar, 0.07 and
0.09 \lsun respectively; GY11 appears much less luminous at $\sim$0.008
\lsun.  These values put all three bodies below the brown dwarf boundary at
$\sim$80 \mj; indeed, GY11 seems to lie close to the deuterium-burning limit
of $\sim$12 \mj (as also argued by Testi et al. 2002).

For comparison, we note that Luhman \& Rieke (1999; hereafter LR99) have
also investigated the properties of $\rho$ Oph members, using $K$-band
spectra and an analysis technique differing from that of NTC02.  For GY310,
their inferred \lbol and \teff, and hence mass from theoretical tracks, are
almost exactly those derived by NTC02.  The brown dwarf status of this
object thus seems secure.  For GY5, their \teff and \lbol (2900K and 0.11
\lsun) are slightly higher than NTC02's values, implying a somewhat larger
mass of $\sim$ 80 \mj.  We cannot be confident therefore that this is a
bona-fide brown dwarf; nevertheless, this higher mass estimate still implies
an object very close to the substellar boundary.  Moreover, using LR99's
estimates does not significantly alter our disk models for this object or
affect our conclusions regarding its disk properties.

The situation is somewhat more complicated for GY11.  LR99 find this object
to be simultaneously much hotter (\teff $\sim$ 2800K) and fainter (\lbol
$\sim$ 0.003 \lsun) than NTC02 derive.  Results similar to LR99's are
obtained by Wilking, Greene \& Meyer (1999; hereafter WGM99) as well; like
LR99, WGM99 also employ K-band spectra.  Theoretical evolutionary tracks
(e.g., D'Antona \& Mazzitelli 1997; Chabrier et al. 2000) then imply a
substantially higher mass, 30--40 \mj, than NTC02's value.  However, they
also then imply an age of $\sim$10 Myr (see comparisons to evolutionary
tracks in LR99 and WGM99).  Now, the infrared excess and veiling of GY11,
signifying large amounts of circum-(sub)stellar material, confirm it to be a
true member of the young $\rho$ Oph star-forming region (WGM99), which has
an estimated age $\lesssim$ 1 Myr.  Thus, the LR99 (and WGM99) values for
GY11 appear suspect.  As WGM99 point out, it is quite possible that veiling
by dust causes its \teff from a K-band analysis to be significantly 
overestimated.  Moreover, the H$_2$O bands used by LR99 and
WMG99 possibly saturate with later M type, which can also result in a
spuriously high \teff determination (LR99).  Spectral type and \teff errors
will lead to erroneous inferred luminosity as well.  The estimates by NTC02
are less susceptible to these particular uncertainties: their $J$- and
$H$-band analysis of the overall spectral shape depends less on H$_2$O
band-strengths alone, and is plausibly less affected by excess emission from
dusty disks than K-band studies.  These considerations, and the fact that
NTC02's luminosity and \teff are in fact consistent with an age $\lesssim$ 1
Myr, suggest that the NTC02 parameters we adopt for GY11 are more reasonable
and appropriate.  In any case, even the higher \teff of LR99 implies a
substellar mass for this object, whether one adopts their \lbol or the NTC02
value ($\sim$ 40 \mj in both cases; also see \S 5).

\section{Disk Models}
To investigate the disk SEDs of our targets, we employ passive disk models
that only consider the effect of reprocessed stellar irradiation and ignore
viscous accretion heating; this is justified for the very low accretion
rates, $\sim$ 10$^{-9}$--10$^{-12}$ {\msun}yr$^{-1}$, inferred so far in the
substellar regime (Muzerolle et al. 2003).  Our models are based on the
2-layer description of Chiang and Goldreich (1997; hereafter CG97) as
modified by Dullemond et al. (2001); scattering is included in an
approximate fashion, following Dullemond \& Natta (2003).  The disk geometry
is assumed to be either flared or flat.  In the former case, consistent with
hydrostatic equilibrium for well-mixed gas and dust
(Kenyon \& Hartmann 1987), the vertical temperature structure at any radius
is described by two components: the temperature in the interior of the disk
and that at the disk surface, which is an optically thin layer directly
exposed to stellar radiation.  The flat, geometrically thin case is modelled
simply as an optically thick blackbody disk which has no silicate emission. 
This situation may correspond to complete settling of large grains to the 
disk midplane (e.g., D'Alessio et al. 2001).  We assume a distance of 150pc 
to the $\rho$ Oph region, consistent with the latest estimate by de Zeeuw et 
al. (1999).

The disks are scaled-down version of typical T Tauri disks; detailed
parameters are given in NTC02 (see also Fig.1 caption).  The gas and dust
are assumed to be well-mixed in the flared models.  As discussed in NTC02,
the emission in the near- and mid-IR depends only on a few disk parameters,
namely the disk inclination and inner radius and the properties of grains on
the disk surface, for which we adopt a mixture of graphite (containg 30\% of 
the cosmic C abundance) and astronomical silicates (containg all cosmic Si 
abundance) with grain radius $a$.  Two distinct particle distributions are
examined: {\it (i)} $a_{min}$ = 0.2$\mu$m and $a_{max}$ = 3.6$\mu$m; and
{\it (ii)} $a_{min}$=3$\mu$m and $a_{max}$=12$\mu$m.  In both cases, the
size distribution goes as d$n$/d$a$ $\propto$ a$^{-3.5}$.  The two
situations allow us to explore the effect of grain growth on the SEDs.  For
the calculations presented here, the surface grains properties are not
crucial as long as $a$ remains smaller than a few microns.  As for stellar
parameters, the most important one is \lbol; for flared disks, there is also
a weak dependence of the flaring angle on the stellar mass and radius
(CG97).

\section{Results and Discussion}
SED model fits to our data for GY310, 11 and 5 are illustrated in Fig. 1.
We plot both our new Subaru data at 8.6, 9.7 and 11.7$\mu$m as well as
additional flux measurements at shorter and longer wavelengths, mainly from
ISO.  The flux-calibrated near-infrared (0.85--2.45$\mu$m) stellar spectra
from NTC02 are also shown.  All the observed fluxes have been corrected for
reddening following NTC02.  For each object, we plot two flared disk models,
one with a distribution of small grains and the other with larger grains; a
flat disk (blackbody) model; and a model stellar spectrum (Allard et al.
2001; see NTC02) depicting the contribution of the underlying stellar
photosphere.  The models incorporate inner holes of various sizes.  The
effect of an inner hole in a flared disk is twofold.  On the one hand, it
depresses the near-IR emission, by getting rid of the innermost hot dust; on
the other hand, it increases the fraction of the stellar surface ``seen'' by
the disk (from 1/2 to 1, in the extreme cases $R_i=R_\ast$ (no hole) and
$R_i\gg R_\ast$ respectively), and therefore the emission at longer
wavelengths.  All three targets exhibit considerable mid-IR excess above the
stellar continuum: while their earlier ISO detections left open the
possibility of source confusion due to ISO's coarse spatial resolution, the
Subaru detections confirm that these objects harbor circum-(sub)stellar
disks.
The salient results from our present work are as follows: {\it (1)} both
GY310 and GY11 clearly possess flared disks - flat disks can be confidently
ruled out in their case, {\it (2)} the 9.7$\mu$m flux for GY310 indicates
possible silicate emission from small surface grains (though larger grains
with no silicate emission cannot be completely excluded), and {\it (3)}
inner holes
of size $\sim$ 3--7 stellar radii seem necessary to fit the data, analogous
to the situation in classical T Tauri stars.  We now briefly discuss each
source.

{\it GY 310}: Our best fit for this object requires a flared disk, tilted
60$^{\circ}$ from face-on and with an inner-hole radius of $R_i$ $\approx$
7$R_{\ast}$.  Inner holes significantly smaller than this produce too much
emission at 6.7$\mu$m, while smaller inclinations overestimate the mid-IR
flux.  Notice that a flat disk model is completely inconsistent with the
data longward of 6.7$\mu$m; the predicted fluxes are simply too low.
Moreover, all the fluxes from 6.7 to 14.3 $\mu$m agree remarkably well with
the flared disk SED that includes 10$\mu$m silicate emission from small
surface grains.  A silicate feature is routinely observed in disks around T
Tauri stars (e.g., Cohen \& Witteborn 1985; Natta, Meyer \& Beckwith 2000).
If our detection is confirmed, it would be the first evidence of such
emission from a brown dwarf disk. Nevertheless, we caution that our present
data cannot unequivocally rule out a flared disk with larger grains and no
silicate feature: our 9.7$\mu$m flux remains marginally consistent with the
latter scenario.  Measurements with smaller uncertainties, such as Spitzer
low-resolution spectra, can settle this issue conclusively.

{\it GY 11}: Our best-fit model for this brown dwarf is a flared disk seen
face-on, with an inner hole of size $R_i$ $\approx$ 3$R_{\ast}$.  Much
smaller inner holes produce unacceptably low mid-IR fluxes.  Similarly, a
flat disk produces emission wholly inadequate to match the steeply rising
mid-IR SED.  However, in the absence of 9.7$\mu$m data, we are not sensitive
to the presence of 10$\mu$m silicate emission, and thus cannot distinguish
between flared disks with small surface grains and those with somewhat
larger particles.  We note that the \lbol we adopt for GY11 (0.012 \lsun),
to obtain an adequate SED fit, is moderately higher than derived by NTC02
(0.008 \lsun).  The NTC02 value produces mid-IR fluxes somewhat lower than
observed (fit not shown); since we already maximize the mid-IR emission by
adopting a face-on disk and invoking an inner hole$\sim$
3$R_{\ast}$, higher luminosity seems the only remaining option in the
context of our models.  The \lbol adopted here is admissible within NTC02's
\av errors for this heavily reddened object (our \lbol amounts to \av
$\approx$8.5 instead of NTC02's 7.5$\pm 1$).  The \teff from synthetic
spectral fits assuming our new \av is 2500K, very close to NTC02's 2400K;
the new \lbol and \teff also remain consistent with an age $\lesssim$ 1 Myr.
  The mass now inferred from the evolutionary tracks is $\sim$ 20 \mj; while
higher than NTC02's estimate of 10 \mj, it still implies that GY11 is a very
low-mass brown dwarf.

{\it GY 5}:   The large scatter in the GY 5 data make any fit provisional at
best.  However, if we disregard the ISO 3.6$\mu$m flux, which is 50\% larger
than the ground-based value in the comparable L$'$ filter (from Comeron et
al. 1998 and Jayawardhana et al. 2003) and assume that the true fluxes at
6.7 and 14.3$\mu$m are at the lower end permissible by the ISO error bars,
then a flat disk inclined at $\sim$ 65$^{\circ}$ to face-on, with an inner
hole $R_i$ $\sim$ 5$R_{\ast}$, provides a plausible fit (especially since
our 9.7$\mu$m Subaru estimate is only a flux upper limit). However, we
cannot discount a flared disk seen at large inclination.

The disk parameters presented here for the three $\rho$ Oph sources are not
unique: there are plenty of trade-offs between factors such as disk
inclination, inner-hole size, grain albedos and sizes, and degree of dust
settling.  For example, our models adopt only the two extremes of grain
settling: well-mixed gas and dust leading to full disk flaring and complete
settling resulting in flat disks.  If there is some, but not complete, dust
settling in GY310 for instance, its moderate mid-IR excess may be fit by a
disk model with a lower inclination.

Nonetheless, the SEDs we observe strongly suggest certain qualitative
features that are relatively independent of the particular disk parameters
adopted.  First, the presence of flared disks in GY310 and 11 is
indubitable: no flat disk, whatever its inner-hole radius or inclination,
can reproduce the observed increase in their flux from 6.7 to 14.3 $\mu$m.
Apai et al. (2002) carried out observations similar to ours for a single
young object at the stellar/substellar edge, ChaH$\alpha$2, and found that a
flat disk with no silicate feature provides the best fit to its SED.  Our
results indicate that, while such disks might indeed be present in some
cases (e.g., perhaps GY5), this does not signify a fundamental change in
disk properties upon crossing from the stellar to the substellar regime:
brown dwarfs can have flared disks too, as well as (tentatively) silicate
emission, and simply present a range of disk geometries just as T Tauri 
stars
do (e.g., Miyake \& Nakagawa 1995).  Second, inner holes of a few substellar
radii may be common in brown dwarf disks - our best fits to all three
objects require such a configuration.  Especially for GY11, it is hard to
find an alternative; even assuming a face-on disk and a (sub)stellar
luminosity higher than previously estimated, some mechanism is required to
further raise the mid-IR fluxes to match the data, and a central hole seems
the only viable possibility (see also Liu et al. 2003).  Indeed,
magnetospheric accretion models predict the formation of inner holes of just
these sizes in both classical T Tauri stars and accreting brown dwarfs, due
to disk truncation by interaction with the (sub)stellar magnetic field
(Shu et al. 1994; Muzerolle et al. 2003).  In
summary, therefore, our study suggests that brown dwarf disk properties are
qualitatively similar to those of higher mass T Tauri stars.  A more
detailed understanding of grain characteristics, dust settling and geometry
in brown dwarf disks will be possible with mid-IR Spitzer observations.

\acknowledgements
We thank the Subaru staff for outstanding support, and Karl Haisch Jr. for
advice on IR reductions.  SM gratefully acknowledges support from the
SIM-YSO grant.  This work was supported in part by NSF grant AST-0205130 to
RJ.  AN acknowledges support from the MIUR grant ``{\it Low-mass stars and
brown dwarfs}''.  MT acknowledges support by Grant-in-Aid (12309010).  DByN
is indebted to the Spanish ``Programa Ram\'on y Cajal'' and AYA2003-05355.
We thank the referees, Ilaria Pascucci and Daniel Apai, for a prompt and 
very helpful review.

\begin{deluxetable}{lccccccccccc}
\tablecaption{\label{tab1} Stellar Parameters and Mid-Infrared Fluxes}
\tablewidth{0pt}
\tablehead{
\colhead{name} &
\colhead{SpT\tablenotemark{a}} &
\colhead{\teff\tablenotemark{a}} &
\colhead{\lbol\tablenotemark{a}} &
\colhead{M\tablenotemark{a}} &
\colhead{\av\tablenotemark{a}} &
\colhead{6.7$\mu$m\tablenotemark{b}} &
\colhead{8.6$\mu$m\tablenotemark{c}} &
\colhead{9.7$\mu$m\tablenotemark{c}} &
\colhead{11.7$\mu$m\tablenotemark{c}} &
\colhead{14.3$\mu$m\tablenotemark{b}} \\
& & (K) & (\lsun) & (\mj) & (mag) & (mJy) & (mJy) & (mJy) & (mJy) & (mJy)\\
}

\startdata

GY5   & M6   & 2700 & 0.07  & 60 & 3.0 & 28$\pm$7 & 19$\pm$3 & $<$35 &
24$\pm$4 & 33$\pm$14 \\
GY11  & M8.5 & 2400 & 0.008 & 10 & 7.0 & 9$\pm$3 & 11$\pm$3 & -- & 21$\pm$5
& 17$\pm$3 \\
& ... & 2500 & 0.012 & 20 & 8.5 &... & ... & ... & ...  & ... \\
GY310 & M6   & 2700 & 0.09  & 60 & 6.0 & 20$\pm$2 & 24$\pm$4 & 35$\pm$5 &
39$\pm$6 & 39$\pm$6 \\

\enddata
\tablenotetext{a}{Spectral type, \teff, \lbol, mass and \av from NTC02.  For
GY11, we also provide, in the second row, the moderately different
parameters suggested by our SED fits.}
\tablenotetext{b}{6.7 and 14.3$\mu$m ISO fluxes from Bontemps et al.
(2001).}
\tablenotetext{c}{8.6, 9.7 and 11.7 $\mu$m fluxes observed by us at Subaru.}

\end{deluxetable}

\plotone{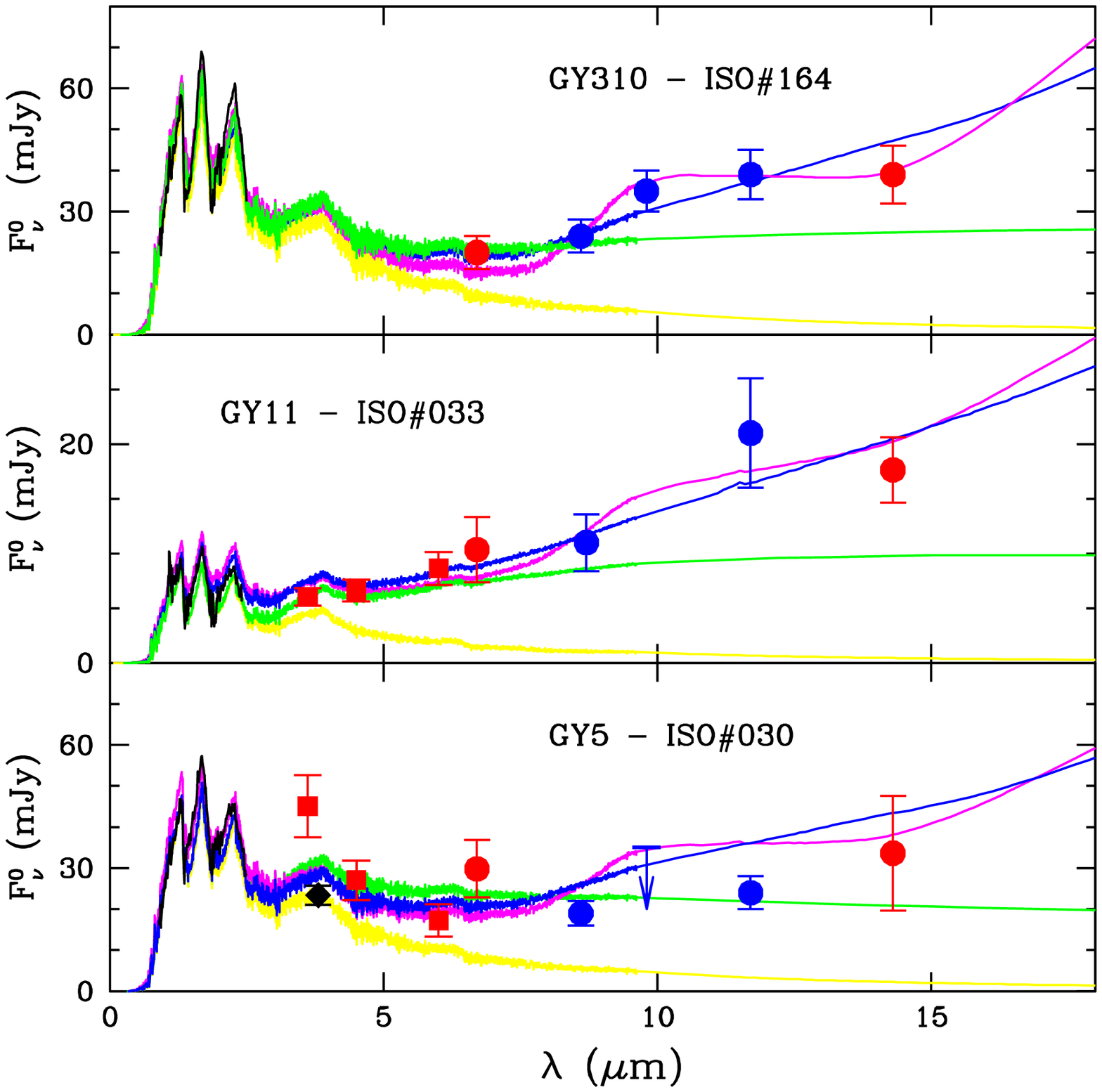}
\figcaption{{\it Top panel}:  Observed fluxes and model predictions for
GY310 
(ISO\#164). Blue dots from this paper, red dots from Bontemps et al. (2000);
black curve shows the near-IR low-resolution spectrum from NTC02.  All
observations are corrected for reddening (\av=6 mag; NTC02). The magenta
curve shows the prediction of a star + flared disk model with small grains;
the blue curve a star + flared disk with large grains; the green curve is
the SED of a flat blackbody disk. The photospheric flux for the adopted
substellar parameters is in yellow. The flared disks have $R_i \sim7R_\ast$,
inclination of 60$^{\circ}$. The flat disk has $R_i \sim 5R_\ast$, same
inclination.   {\it Middle Panel}: Same for GY11 (ISO\#033). The red squares
are ISO fluxes from Comer\'on et al. (1998).  The adopted stellar
parameters (see \S 5) are \lbol=0.012\lsun, \av=8.5 mag. All disks have
$R_i\sim 3R_\ast$ and are seen face-on.  {\it Bottom Panel}: Same for GY5
(ISO\#030).  The black diamond is an L$'$ flux ground-based measurement from
Comer\'on et al. (1998).  The flared disks have $R_i \sim 5R_\ast$,
inclination $\sim 60^{\circ}$. The flat disk has $R_i \sim 5R_\ast$ and
inclination $\sim 65^{\circ}$.  For all 3 objects, the other disk 
parameters are: disk outer 
radius of $R_D$ = 1$\times$10$^{15}$ cm (67 AU); disk mass of $M_D$ = 0.03
$M_\ast$; disk surface density profile ${\Sigma}\,{\propto}\,R^{-1}$.  The
dust in the disk interior is assumed to have an opacity $\kappa$ =
0.01($\lambda$/1.3 mm)$^{-1}$ cm$^2$g$^{-1}$ (Beckwith et al. 1990). Overall
results in this figure do not depend on the exact values of these
parameters.}

\end{document}